\documentclass[twocolumn,showpacs,preprintnumbers,amsmath,amssymb,superscriptaddress]{revtex4}

\usepackage{graphicx,epsfig}
\usepackage{dcolumn}
\usepackage{bm}

\begin{document}

\title{ Search for the decay $K^+\to \pi^+ \nu \bar\nu$ in the 
momentum region $P_\pi < 195 {\rm ~MeV/c}$ }

\author{S.~Adler} \affiliation{Brookhaven National Laboratory, Upton,
New York 11973} 

\author{M.~Aoki}
\altaffiliation{Present address: Department of Physics, Osaka
University, Toyonaka, Osaka 560-0043, Japan.} 
\affiliation{TRIUMF, 4004 Wesbrook Mall, Vancouver, British Columbia,
Canada, V6T 2A3} 

\author{M.~Ardebili} \affiliation{Joseph Henry
Laboratories, Princeton University, Princeton, New Jersey 08544}

\author{M.S.~Atiya} \affiliation{Brookhaven National Laboratory, Upton,
New York 11973} 

\author{A.O.~Bazarko} \affiliation{Joseph Henry
Laboratories, Princeton University, Princeton, New Jersey 08544}

\author{P.C.~Bergbusch} \affiliation{Department of Physics and
Astronomy, University of British Columbia, Vancouver, British
Columbia, Canada, V6T 1Z1} 

\author{B.~Bhuyan} \altaffiliation{Also at 
Department of Physics and Astrophysics, 
University of Delhi, Delhi, India}  
\affiliation{Brookhaven National Laboratory, Upton,
New York 11973} 

\author{E.W.~Blackmore}
\affiliation{TRIUMF, 4004 Wesbrook Mall, Vancouver, British Columbia,
Canada, V6T 2A3} 

\author{D.A.~Bryman} \affiliation{Department of
Physics and Astronomy, University of British Columbia, Vancouver,
British Columbia, Canada, V6T 1Z1}

\author{I-H.~Chiang} \affiliation{Brookhaven National
Laboratory, Upton, New York 11973}

\author{M.R.~Convery} \affiliation{Joseph Henry
Laboratories, Princeton University, Princeton, New Jersey 08544}

\author{M.V.~Diwan}
\affiliation{Brookhaven National Laboratory, Upton, New York 11973}

\author{J.S.~Frank} \affiliation{Brookhaven National Laboratory,
Upton, New York 11973} 

\author{J.S.~Haggerty} \affiliation{Brookhaven
National Laboratory, Upton, New York 11973}

\author{T.~Inagaki} \affiliation{High Energy
Accelerator Research Organization (KEK), Oho, Tsukuba, Ibaraki
305-0801, Japan} 

\author{M.~Ito}\altaffiliation{Present address:
Thomas Jefferson
National Accelerator Facility, Newport News, Virginia 23606.}
\affiliation{Joseph Henry Laboratories,
Princeton University, Princeton, New Jersey 08544}

\author{V.~Jain}
\affiliation{Brookhaven National Laboratory, Upton, New York 11973}

\author{D.E.~Jaffe}
\affiliation{Brookhaven National Laboratory, Upton, New York 11973}

\author{S.~Kabe} \affiliation{High Energy Accelerator Research
Organization (KEK), Oho, Tsukuba, Ibaraki 305-0801, Japan}

\author{M.~Kazumori} \affiliation{High Energy
Accelerator Research Organization (KEK), Oho, Tsukuba, Ibaraki
305-0801, Japan} 

\author{S.H.~Kettell} \affiliation{Brookhaven National Laboratory,
Upton, New York 11973}
 
\author{P.~Kitching} \affiliation{Centre for
Subatomic Research, University of Alberta, Edmonton, Canada, T6G 2N5}

\author{M.~Kobayashi} \affiliation{High Energy Accelerator Research
Organization (KEK), Oho, Tsukuba, Ibaraki 305-0801, Japan}

\author{T.K.~Komatsubara} \affiliation{High Energy Accelerator
Research Organization (KEK), Oho, Tsukuba, Ibaraki 305-0801, Japan}

\author{A.~Konaka} \affiliation{TRIUMF, 4004 Wesbrook Mall, Vancouver,
British Columbia, Canada, V6T 2A3}

\author{Y.~Kuno}
\altaffiliation{Present address: Department of Physics, Osaka
University, Toyonaka, Osaka 560-0043, Japan.} 

 \affiliation{High
Energy Accelerator Research Organization (KEK), Oho, Tsukuba, Ibaraki
305-0801, Japan} 

\author{M.~Kuriki} 
\affiliation{High Energy
Accelerator Research Organization (KEK), Oho, Tsukuba, Ibaraki
305-0801, Japan} 

\author{T.F.~Kycia} \altaffiliation{Deceased}\affiliation{Brookhaven 
National Laboratory, Upton, New York 11973}

\author{K.K.~Li} \affiliation{Brookhaven National Laboratory,
Upton, New York 11973}

\author{L.S.~Littenberg} \affiliation{Brookhaven National Laboratory,
Upton, New York 11973} 

\author{D.R.~Marlow} \affiliation{Joseph Henry
Laboratories, Princeton University, Princeton, New Jersey 08544}

\author{R.A.~McPherson} \affiliation{Joseph
Henry Laboratories, Princeton University, Princeton, New Jersey 08544}

\author{J.A.~Macdonald} \affiliation{TRIUMF,
4004 Wesbrook Mall, Vancouver, British Columbia, Canada, V6T 2A3}

\author{ P.D.~Meyers} \affiliation{Joseph Henry Laboratories,
Princeton University, Princeton, New Jersey 08544}

\author{J.~Mildenberger} \affiliation{TRIUMF, 4004 Wesbrook Mall,
Vancouver, British Columbia, Canada, V6T 2A3}

\author{N.~Muramatsu}
\altaffiliation{Present address: Japan Atomic Energy Research
Institute, Sayo, Hyogo 679-5198, Japan.}
\affiliation{High Energy Accelerator Research Organization (KEK), Oho,
Tsukuba, Ibaraki 305-0801, Japan}

\author{T.~Nakano}
\affiliation{Research Center for Nuclear Physics, Osaka University,
10-1 Mihogaoka, Ibaraki, Osaka 567-0047, Japan} 

\author{C.~Ng}
\altaffiliation{Also at Physics Department, State University of New
York at Stony Brook, Stony Brook, NY 11794-3800.}
\affiliation{Brookhaven National Laboratory, Upton, New York 11973}

\author{S.~Ng} \affiliation{Centre for
Subatomic Research, University of Alberta, Edmonton, Canada, T6G 2N5}

\author{T.~Numao} \affiliation{TRIUMF, 4004 Wesbrook Mall, Vancouver,
British Columbia, Canada, V6T 2A3}

\author{J.-M.~Poutissou}
\affiliation{TRIUMF, 4004 Wesbrook Mall, Vancouver, British Columbia,
Canada, V6T 2A3} 

\author{R.~Poutissou} \affiliation{TRIUMF, 4004
Wesbrook Mall, Vancouver, British Columbia, Canada, V6T 2A3}

\author{G.~Redlinger} \altaffiliation{Present address: Brookhaven National Laboratory.}
\affiliation{TRIUMF, 4004 Wesbrook Mall, Vancouver, British Columbia,
Canada, V6T 2A3} 

\author{T.~Sasaki}
\affiliation{Research Center for Nuclear Physics, Osaka University,
10-1 Mihogaoka, Ibaraki, Osaka 567-0047, Japan} 

\author{T.~Sato} \affiliation{High Energy Accelerator
Research Organization (KEK), Oho, Tsukuba, Ibaraki 305-0801, Japan}

\author{T.~Shinkawa} \altaffiliation{Present address: National Defense
Academy of Japan, Yokosuka, Kanagawa 239-8686, Japan.}
\affiliation{High Energy Accelerator Research Organization (KEK), Oho,
Tsukuba, Ibaraki 305-0801, Japan}

\author{F.C.~Shoemaker}
\affiliation{Joseph Henry Laboratories, Princeton University,
Princeton, New Jersey 08544} 

\author{A.J.S.~Smith}
\affiliation{Joseph Henry Laboratories, Princeton University,
Princeton, New Jersey 08544} 

\author{R. Soluk} \affiliation{Centre for
Subatomic Research, University of Alberta, Edmonton, Canada, T6G 2N5}

\author{J.R.~Stone} \affiliation{Joseph
Henry Laboratories, Princeton University, Princeton, New Jersey 08544}

\author{R.C.~Strand} \affiliation{Brookhaven National Laboratory,
Upton, New York 11973}

\author{S.~Sugimoto} \affiliation{High Energy
Accelerator Research Organization (KEK), Oho, Tsukuba, Ibaraki
305-0801, Japan} 

\author{Y.~Yoshimura}
\affiliation{High Energy Accelerator Research Organization (KEK), Oho,
Tsukuba, Ibaraki 305-0801, Japan}

\author{C.~Witzig} \affiliation{Brookhaven National
Laboratory, Upton, New York 11973}

\collaboration{E787 Collaboration}

\date{\today}

\begin{abstract}
We have searched for the decay $K^+ \to \pi^+ \nu \bar\nu$ in the 
kinematic region with pion momentum below the $K^+ \to \pi^+ \pi^0$ peak. 
One event was observed, 
consistent with the background estimate of $0.73\pm 0.18$.
This implies an upper limit 
on $B(K^+ \to \pi^+ \nu \bar\nu)< 4.2\times 10^{-9}$ (90\% C.L.), 
  consistent with the recently measured 
branching ratio of $(1.57^{+1.75}_{-0.82})
\times 10^{-10}$,  obtained using the standard model spectrum and 
the kinematic region  above the
$K^+ \to \pi^+ \pi^0$ peak. 
The same data were  used to search for 
 $K^+ \to \pi^+ X^0$,
where $X^0$ is a weakly interacting neutral particle or system
of particles with $150 < M_{X^0} < 250 {\rm ~MeV/c^2}$. 
\end{abstract}

\maketitle

In a recent paper  we reported   the branching ratio  for 
the rare decay $K^+\to \pi^+ \nu \bar\nu$ to be 
$(1.57^{+1.75}_{-0.82})\times 10^{-10}$ based on the observation of two
events in the phase 
space region $p_{\pi^+}> 211 {\rm ~MeV/c}$ \cite{pnn1}. 
This decay is sensitive  to the coupling of top to down 
quarks, $V_{td}$, in the Cabibbo-Kobayashi-Maskawa  mixing matrix.
The standard model (SM) predicted branching 
ratio, $B(K^+ \to \pi^+ \nu \bar\nu)$,  is  $(0.75 \pm 0.29) \times 10^{-10}$\cite{smp}. 
Loop diagrams involving new heavy particles in extensions of 
 the SM can interfere with SM diagrams
and alter the decay rate, and also  the 
kinematic  spectrum\cite{bsmp}. Exotic scenarios such 
$K^+ \to \pi^+ X^0$ where $X^0$ is a hypothetical 
stable weakly interacting particle or system of particles have also been 
suggested \cite{bs, aliev}.
 It is therefore 
important 
to obtain  higher statistics for this decay and to extend the measurement 
to other regions of phase space.  
The  results in \cite{pnn1} are from analysis of data 
with the $\pi^+$ momentum 
above the $K^+\to \pi^+ \pi^0$ ($K_{\pi 2 }$) peak (Region 1). 
The $\pi^+$ from $K_{\pi2}$
decay has a kinetic energy ($E$), momentum ($P$), and range ($R$)
 in plastic scintillator of 
108 MeV, 205 MeV/c, and 30 cm, respectively. 
In this letter we report the analysis of 
 data below the $K_{\pi 2}$ peak (Region 2) obtained  from 
Experiment E787\cite{nim1,ccd,ec,utc,td,pnn197,pnn195}
at the Alternating 
Gradient Synchrotron  (AGS) of Brookhaven National Laboratory. 
  The previous limit 
using Region 2, $1.7\times 10^{-8}$ (90\% C.L.), 
 was obtained  from an earlier version  of  
the E787 detector\cite{pnn2early}.

The signature for $K^+\to \pi^+ \nu \bar\nu$ 
in the E787 experiment 
is a single  $K^+$ stopping in a   target (TG),
decaying 
 to a single $\pi^+$ with no other accompanying photons or charged particles.  
In Region 1, the major backgrounds were found  to be 
the two body decays  $K_{\pi2}$
and  $K^+ \to \mu^+ \nu_\mu ~(K_{\mu2})$, 
scattered beam pions, 
and $K^+$ charge exchange (CEX) reactions resulting in 
decays $K^0_L \to \pi^+ l^- \bar\nu_l$, where $l = e$ or $\mu$. 
Region 2 has larger potential acceptance than Region 1 
because the  phase space is more than twice as large and the 
loss of pions due to nuclear interactions in the detector is smaller 
at the lower pion energies. However, there are additional sources of 
background for Region 2. These include $K_{\pi 2}$ in which the $\pi^+$ 
loses energy by scattering 
 in the material of the detector (primarily in the TG), $K^+ \to \pi^+ \pi^0 \gamma ~(K_{\pi2\gamma})$,
$K^+ \to \mu^+ \nu \gamma ~(K_{\mu2\gamma})$, 
$K^+ \to \mu^+ \nu \pi^0 ~(K_{\mu3})$,   
and
 $K^+ \to \pi^+ \pi^- e^+ \nu_e $ ($K_{e4}$) 
decays in which both the $\pi^-$ and the $e^+$ are 
invisible because of absorption.

The data were obtained with a flux of 
$6\times 10^6$ kaons per 1.6 sec spill at  730 MeV/c  (with 24\% pion contamination)
  entering the apparatus.
The kaons were identified  by a Cerenkov detector; 
two multi-wire-proportional-chambers were used to determine 
that there was only one entering particle. 
After slowing in a  BeO degrader  the kaons  traversed a 
10-cm-thick lead-glass detector read out by 16 fine-mesh
photomultiplier tubes (PMT) and a scintillating  target hodoscope (TH)
placed before  the TG.
 The lead-glass detector  was designed to 
be insensitive to kaons and  
detect  electromagnetic showers  originating 
from kaon decays in the TG. 
The TH  was used to verify that there was only one kaon 
as well as determine 
the  position, time, and energy loss of the kaon before 
it entered and stopped in the TG. 
The TG  consisted of 413 5.0-mm-square, 
3.1-m-long plastic scintillating fibers, each connected to a PMT. 
The fibers were packed axially to form a cylinder of $\sim$12 cm  diameter.
Gaps in the outer edges of the TG were filled with smaller fibers 
which were connected to PMTs in groups. 
The PMTs were read out by ADCs, TDCs, and 500 MHz transient digitizers based on GaAs charge-coupled 
devices (CCDs)\cite{ccd}. 
Photons  were detected in a hermetic calorimeter mainly 
 consisting of a 14-radiation length thick barrel detector made of 
lead/scintillator sandwich and 13.5-radiation length thick 
endcaps of undoped CsI crystals \cite{ec}.
The rest of the detector 
consisted of a central drift chamber (UTC)\cite{utc}, 
and a cylindrical 
range stack (RS)   of 21 layers of plastic 
scintillator with two layers of embedded tracking chambers, 
all within a 1-T solenoidal magnetic field. The TG, UTC, and RS 
allowed the measurement of the $P$, $R$, and $E$ 
of the charged decay products.  The $\pi \to \mu \to  e$ 
decay sequence from pions that came to rest in the RS was observed using 
another set of 500 MHz transient digitizers (TD)\cite{td}.

\begin{figure}
\begin{center}
\vspace{-2cm}
\epsfig{figure=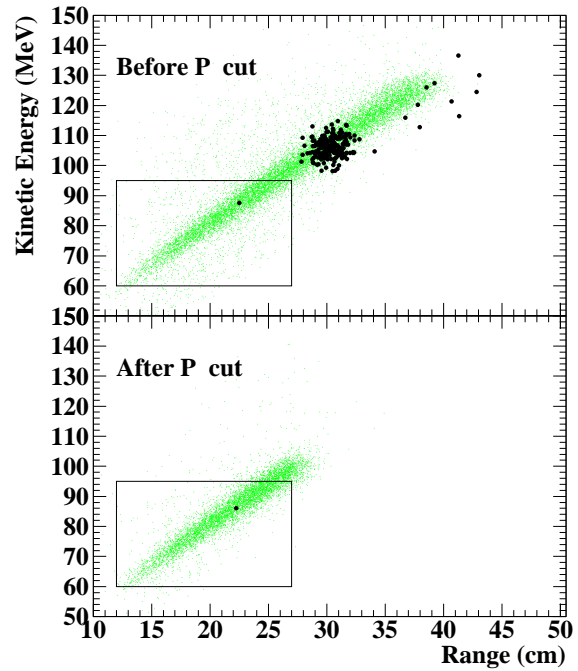, width=4.5in}
\vspace{-1.25cm} 
\end{center}
\caption{ Range (cm in plastic scintillator) and kinetic 
energy (MeV) of events remaining after all cuts 
except the momentum cut (top), and including the momentum cut (bottom).
 The dark points 
represent  the data. The simulated distribution of 
expected events from $K^+ \to \pi^+ \nu \bar\nu$
is indicated by the light dots.  The group of events around 108 MeV is due to the 
$K_{\pi2}$ background. The  events at higher energy are due
 to $K_{\mu2}$ and $K_{\mu2\gamma}$  background. 
All events except for the one in the signal box 
 are  eliminated by the $140 < P <195 {\rm ~MeV/c}$ 
cut on momentum. 
\vspace{-0.5cm} 
}
\label{rpeplt}
\end{figure}

The data reduction and offline  analysis for Region 2  was 
similar to the analysis of Region  1 \cite{bergbush,pnn197,pnn195},
 although the final cuts 
to enhance signal and  suppress background to less than 
one event  were different.  
 Here
we will emphasize the key instrumentation and 
analysis tools  used to suppress the background in Region 2. 
 The TG, CCDs, and the photon veto system
were  the important elements for Region 2 analysis. 
A multilevel trigger selected events 
by requiring 
 an identified $K^+$ to stop 
in the TG, followed,
after a delay of at least 1.5 ns, by a single charged particle track that
traversed TG and 
 RS  with a hit-pattern  
consistent with the expectation for $K^+ \to \pi^+ \nu \bar\nu$.
Events with photons were suppressed  by vetos on the barrel 
and endcap detectors. 
In the ~~offline analysis,
the single charged particle 
 was required to be identified as a $\pi^+$ with 
$P$, $R$, and $E$ consistent for a $\pi^+$, and 
the TD pulse information consistent with the decay sequence 
$\pi \to \mu \to e$  in  the 
last RS counter on the pion trajectory.
The signal region was defined by the intervals $140 < P < 195 {\rm ~MeV/c}$,
$12 < R < 27 {\rm ~cm}$, and $60 < E < 95 {\rm ~MeV}$. 

The background was found to be dominated by 
$K_{\pi 2}$ events in which the pion had a  
nuclear interaction near the kaon decay vertex, most probably 
on a carbon nucleus in 
the TG plastic scintillator. This scatter left
 the pion with  reduced 
kinetic energy, putting it in Region 2. 
We suppressed
 this background by removing events in which the pion track 
had a scattering signature in the TG. These signatures included 
kinks, tracks  that did not  point back to  the vertex fiber in which the 
kaon decayed,
 or energy deposits inconsistent with the 
ionization energy loss for a pion 
of the measured momentum. 
The remaining $K_{\pi2} $ background consisted of events in 
which the pion scattered  in one of the fibers 
 traversed by the kaon.  The extra energy deposits from the
pion scatters were obscured by the earlier large energy 
deposits of the kaon. For these events, 
we examined the  pulse shapes recorded in  the CCDs in each
 kaon  fiber  using a $\chi^2$ fit and eliminated  events in which   
an overlapping 
second pulse, in time with the pion, was found to have energy larger
 than 1 MeV. 
To obtain sufficient separation of the $K^+$  and $\pi^+$ induced pulses 
in the CCDs we  required a minimum
delay of 6 ns between the kaon and the pion. 
Finally, additional $K_{\pi2}$ rejection was obtained by removing 
events 
with photon interactions in detectors surrounding the kaon beam-line;
these cuts caused substantial ($\sim$ 42 \%) 
loss of efficiency because of accidental hits due to the high
flux of particles. 


\begin{table} 
\begin{center}
\begin{tabular}{|l|l|r|}
\hline 
$K^+ \to \pi^+ \pi^0$ & d    &  $0.630\pm0.170$   \\
$K^+ \to \pi^+ \pi^0 \gamma$ & dm  &  $0.027\pm 0.004$  \\
$K_{\mu2\gamma} + K_{\mu3}$ & d &  $0.007\pm 0.007$  \\ 
Beam   & d &   $0.033\pm 0.033$   \\
$K^+ \to \pi^+ \pi^- e^+ \nu_e$ & dm &   $0.026\pm 0.032$       \\
CEX & dm &   $0.011\pm 0.011$    \\
\hline 
Total  &   &   $0.734\pm 0.177$    \\
\hline 
\end{tabular}
\end{center}
\caption{Estimated number of background events  
for  Region 2 of $K^+\to \pi^+ \nu \bar\nu$ data.
The second column indicates the method of background determination:
 data alone (d), data combined with simulation (dm).   
The errors include statistics of the data and  Monte Carlo as well as  
systematic uncertainties. \vspace{-0.5cm} }
\label{tab1}
\end{table}

We formed 
multiple  independent constraints  on each source of 
background. These constraints were grouped in two independent sets of cuts, 
designed to have  little  correlation.  One 
set of cuts was relaxed (or inverted) to enhance the background so that 
the other set could be evaluated to determine its power of rejection, as 
summarized below.
The sum of the background due to  
$K_{\mu 2\gamma}$ and $K_{\mu 3}$  was
obtained 
 by separately measuring the rejection factors 
 of the TD particle identification 
and kinematic ($R$ and $P$) particle identification. The background 
from beam pions  was evaluated by 
separately measuring the rejections of Cerenkov,
  TH  beam particle identification, 
and the delay time between pion and kaon. 
The dominant background from $K_{\pi2}$ decay was measured by 
evaluating the rejection of the photon veto system on events tagged by
scattering signatures in the TG and target CCDs.
Similarly, the rejection of the target CCD cut was determined
 by using events 
that failed  the photon veto criteria.
It should be noted that  the Region 1 analysis measured 
 photon veto rejection using the unscattered events in 
the momentum peak (205 MeV/c)\cite{pnn195}. 
This method could not be used for Region 2 because the 
scattering in the TG
spoiled  the back-to-back correlation 
between the detected $\pi^+$ track in the RS and the undetected 
$\pi^0$, leading to different photon veto rejection factors  for scattered
and unscattered $K_{\pi2}$ background events.

We employed Monte Carlo (MC) simulation
to evaluate the backgrounds from 
$K_{e4}$ and $K_{\pi2\gamma}$ 
because these could not be distinguished,  
on the basis of the $\pi^+$ track alone,
from the much larger $K_{\pi2}$ background with a $\pi^+$ scatter.  
In the case of   $K_{e4}$, we 
first identified such events by looking
 for additional short tracks, indicative of  $\pi^-$ or $e^+$,  
coming from the kaon decay
vertex in TG. 
 We then used MC events to estimate the power of 
rejecting background 
events with charged particle hits unrelated 
to the $K^+$ or the $\pi^+$ in the TG.
The  MC simulation was performed 
 using the previously measured decay distribution of 
$K_{e4}$ decays\cite{ke4}. For the final  estimate (Table \ref{tab1})
we relied on the observed number of $K_{e4}$ candidate events,
which were found to be consistent with  the known branching ratio. 
For  simulation of $\pi^-$ absorption  we used a measurement 
of ionization spectra 
in scintillator performed previously by our collaboration 
\cite{pimumu} and corroborating 
information from other sources\cite{pimcross,pimspec}. 
For the $K_{\pi2\gamma}$ background, a MC calculation 
provided the ratio between the
 observed number of events 
in the $K_{\pi2}$ peak  and the 
 expected background.   
The calculation was performed 
using the previously 
measured \cite{ppg} decay rate
 and  a calculation of  the 
extra  rejection, based on calibration data,  
due to the radiative photon in  $K_{\pi2 \gamma}$ decays. 
Measurements of $K^+$ charge exchange reaction in the TG were performed.
These 
measurements were used as input to the MC calculation 
of the CEX background. 
 The final background estimates and associated errors in 
Table \ref{tab1} include   
corrections for    
small correlations  in the separate groups of cuts
and cross contamination of background samples.

\begin{table}
\begin{center}
\begin{tabular}{|l|r|}
\hline 
Acceptance factors  &  \\
\hline 
$K^+$ stop efficiency & 0.670  \\
$K^+$ decay after 6 ns & 0.591 \\
$K^+ \to \pi^+ \nu \bar\nu$ phase space &  0.345  \\
Geometry   & 0.317  \\   
$\pi^+$ nucl. int. and decay in flight & 0.708  \\
Reconstruction efficiency &  0.957  \\    
Other kinematic cuts & 0.686   \\     
$\pi - \mu - e$ decay chain & 0.545  \\
Beam and target analysis  & 0.479  \\    
CCD acceptance  &  0.401   \\
Accidental loss &  0.363  \\
\hline 
Total acceptance & $7.65\times 10^{-4}$   \\
\hline 
\end{tabular}
\end{center}
\caption{Acceptance factors used in the measurement of 
$K^+\to \pi^+ \nu \bar\nu$  in Region 2. 
The ``$K^+$ stop efficiency'' is the fraction of kaons entering 
the TG that stopped.
``Other kinematic constraints'' include 
particle identification cuts.\vspace{-0.5cm}}
\label{accept}
\end{table}

\begin{figure}
\begin{center}
\vspace{-0.5cm}
\epsfig{figure=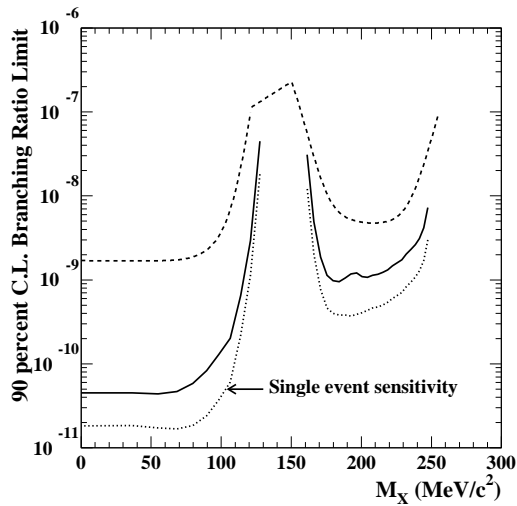, width=3in}
\vspace{-0.5cm}
\end{center}
\caption{ The 90\% C.L. upper limit for $B(K^+ \to \pi^+ X^0$) as a 
function of $M_{X^0}$, the mass of the recoiling system. 
The solid  (dashed) line is from this analysis (from \cite{pnn2early}).  
The limit for $M_{X^0}< 140 {\rm ~MeV/c^2}$ 
is derived from the result for Region 1. 
The observation of one event in Region 2 reported in this Letter
causes a bump in the limit at $194 {\rm ~MeV/c^2}$.
Similarly the  2 events, consistent with the observation of 
$K^+ \to \pi^+ \nu \bar\nu$ above background,  
in Region 1 reported in \cite{pnn1}
increase  the limit at 
$105$ and $86 {\rm ~MeV/c^2}$.
We have also  
included  the single event sensitivity 
as a function of  $M_{X^0}$ (the dotted line), defined in the text, 
obtained by E787.
\vspace{-0.5cm}}
\label{xplot}
\end{figure}

The integrity of the background estimates was assured because 
the background cuts were defined  using only one-third of the data,
sampled uniformly from 
the entire set,
without examining the events in the pre-determined signal region. 
The cuts were then applied with no further changes
to the  remaining two-thirds of the data to obtain the numbers 
reported in Table \ref{tab1}.
The systematic error on the largest background, $K_{\pi2}$, 
was estimated  by measuring the 
rejection of photon veto cuts on many different event 
ensembles, tagged in different ways for a TG scattering signature.
The event ensembles were designed to have little  contamination by
other background sources such as $K_{\pi2\gamma}$ and $K_{e4}$. 
As a final check, 
each cut  was  relaxed to admit background events 
in a predictable way. 
Examination of these background events,
which are close to the signal region, provided no indication of 
background sources other than those in Table \ref{tab1}. 
For example, 
the kaon decay time region between 2 to 6 ns, with acceptance 
of $0.254\pm 0.004$ (less than the naive expectation due to other
lifetime-dependent cuts) relative to the 
signal region,  was examined.
This region has a total estimated 
background of $0.45\pm 0.14$, dominated by $K_{\pi2}$ decays due 
to the reduced background rejection from 
the CCDs. One event, consistent with the background estimate, 
was observed in this background
region.

After the background study, the signal region  was examined, yielding 
one candidate event with $P=180.7 {\rm ~MeV/c}$, $R=22.1 {\rm ~cm}$, and 
$E=86.3 {\rm ~MeV}$ with a kaon 
decay time of 17.7 ns, consistent with the background estimate of $0.73\pm 0.18$. 
Fig.~\ref{rpeplt} shows the 
kinematics of the remaining events  
before and after the cut on measured momentum, $P$. 

Using the total number of $K^+$ incident on TG for these data, 
 $1.12\times 10^{12}$, 
 the acceptance reported in Table \ref{accept}, and  the observation of one 
event in  Region 2  we calculate the
 upper limit of $B(K^+\to \pi^+ \nu \bar\nu) < 4.2\times 10^{-9}$ (90\% C.L.) 
\cite{feldman}. This is  
consistent with the branching ratio 
reported from Region 1 
and the SM decay spectrum \cite{pnn1}; combining the measurements from Region 1 and 
Region 2 does not alter the branching ratio measurement significantly because it is 
dominated by the sensitivity of Region 1. However,  
for non-standard scalar and tensor  interactions,
 Region 2 has  larger  acceptance  than Region 1. 
We have combined
the sensitivity of both regions to obtain 
the 90\% C.L. upper limits, $4.7\times 10^{-9}$ and 
$2.5\times 10^{-9}$, for scalar and tensor interactions, respectively. 

This measurement is also sensitive to $K^+ \to \pi^+ X^0$, 
where $X^0$ is a hypothetical 
stable weakly interacting particle, or system
 of particles. Fig.~\ref{xplot} shows 90\% C.L. upper 
limits on $B(K^+ \to \pi^+ X^0)$ together with
the previous limit from \cite{pnn2early}. 
The dotted line in Fig.~\ref{xplot} is 
the single event sensitivity defined as the
inverse of the acceptance for $K^+ \to \pi^+ X^0$ multiplied by the total 
number of stopped kaons as a
function of  $M_{X^0}$. 

In conclusion, 
the use of GaAs charged-coupled devices to record pulse shapes 
as well as highly efficient photon detection has allowed us to 
suppress background in Region 2. This has resulted in new limits on 
the spectrum of the pion in the decay  $K^+ \to \pi^+ \nu \bar\nu$
as well as improvement in the sensitivity to $K^+ \to \pi^+ X^0$
by a factor between 4 and 40 over the accessible mass range. 
The detailed enumeration of backgrounds in Region 2  will be
important for  new experiments that intend to precisely
measure  $K^+ \to \pi^+ \nu \bar\nu$ 
with large statistics \cite{nim2,ckm}.

We gratefully acknowledge the dedicated efforts of the technical staff 
supporting this experiment and the Brookhaven Accelerator Department. 
This research was supported in part by the U.S. Department of Energy 
under Contracts No. DE-AC0298CH10886, W-7405-ENG-36, and grant 
DE-FG02-91ER40671, by the Ministry of Education, 
Culture, Sports, Science and Technology
of Japan through the Japan-US Cooperative Research 
Program in High Energy Physics and under the Grant-in-Aids for Scientific 
Research, encouragement of Young Scientists and for JSPS Fellows, and by 
the Natural  Sciences and Engineering Research Council and the National 
Research Council of Canada.


\begin{thebibliography}{99}
\bibitem{nim2} An upgraded version of the experiment, E949, is now pursuing 
a precise measurement of $K^+ \to \pi^+ \nu \bar\nu$.
\end{thebibliography}
\end{document}